# Remote-sensing characterization of major Solar System bodies with the Twinkle space telescope

Billy Edwards
Giorgio Savini
Giovanna Tinetti
Marcell Tessenyi
Claudio Arena
Sean Lindsay
Neil Bowles

**SPIE.**

# Remote-sensing characterization of major Solar System bodies with the Twinkle space telescope

Billy Edwards,[a,*] Giorgio Savini,[a,b] Giovanna Tinetti,[a,b] Marcell Tessenyi,[a,b] Claudio Arena,[a] Sean Lindsay,[c] and Neil Bowles[d]
[a]University College London, Department of Physics and Astronomy, London, United Kingdom
[b]Blue Skies Space Ltd., London, United Kingdom
[c]University of Tennessee, Department of Physics and Astronomy, Knoxville, Tennessee, United States
[d]University of Oxford, Atmospheric, Oceanic and Planetary Physics, Clarendon Laboratory, Oxford, United Kingdom

**Abstract.** Remote-sensing observations of Solar System objects with a space telescope offer a key method of understanding celestial bodies and contributing to planetary formation and evolution theories. The capabilities of Twinkle, a space telescope in a low Earth orbit with a 0.45-m mirror, to acquire spectroscopic data of Solar System targets in the visible and infrared are assessed. Twinkle is a general observatory that provides on-demand observations of a wide variety of targets within wavelength ranges that are currently not accessible using other space telescopes or that are accessible only to oversubscribed observatories in the short-term future. We determine the periods for which numerous Solar System objects could be observed and find that Solar System objects are regularly observable. The photon flux of major bodies is determined for comparison to the sensitivity and saturation limits of Twinkle's instrumentation and we find that the satellite's capability varies across the three spectral bands (0.4 to 1, 1.3 to 2.42, and 2.42 to 4.5 $\mu m$). We find that for a number of targets, including the outer planets, their large moons, and bright asteroids, the model created predicts that with short exposure times, high-resolution spectra (R ~ 250, $\lambda$ < 2.42 $\mu m$; R ~ 60, $\lambda$ > 2.42 $\mu m$) could be obtained with signal-to-noise ratio (SNR) of > 100 with exposure times of <300 s. For other targets (e.g., Phobos), an SNR > 10 would be achievable in 300 s (or less) for spectra at Twinkle's native resolution. Fainter or smaller targets (e.g., Pluto) may require multiple observations if resolution or data quality cannot be sacrificed. Objects such as the outer dwarf planet Eris are deemed too small, faint or distant for Twinkle to obtain photometric or spectroscopic data of reasonable quality (SNR > 10) without requiring large amounts of observation time. Despite this, the Solar System is found to be permeated with targets that could be readily observed by Twinkle. © The Authors. Published by SPIE under a Creative Commons Attribution 4.0 Unported License. Distribution or reproduction of this work in whole or in part requires full attribution of the original publication, including its DOI. [DOI: 10.1117/1.JATIS.5.1.014006]

Keywords: space telescope; visible and near-infrared spectroscopy; Solar System bodies.

Paper 18069 received Aug. 28, 2018; accepted for publication Feb. 19, 2019; published online Mar. 15, 2019.

## 1 Introduction

Spacecraft studies of Solar System bodies have increasingly contributed to our knowledge of these objects over recent years. *In-situ* measurements provide the best means of understanding a target, but dedicated lander, orbiting or flyby missions are rare and thus remote-sensing missions offer a great chance to observe an object of interest. Some targets can be viewed from ground-based telescopes at certain wavelengths (e.g., visible), but significant issues are encountered in other bands due to atmospheric absorption, particularly if observing at infrared or UV wavelengths. In addition, ground observations can be affected by weather and atmospheric distortion. Space telescopes avoid these issues and thus are fundamental to increasing our knowledge of planetary science.

Potential targets for observation within our local stellar environment are diverse and each offers insight into the Solar System as a whole. Asteroids and comets are remnants of the earliest celestial bodies, providing a means of investigating the formation of the planets we know today (e.g., Refs. 1–3). Studying the building blocks of the Solar System, as well as the larger bodies to have formed, enhances our understanding of planet formation and evolution. Spectroscopic observations, particularly at visible and infrared wavelengths, allow the composition of the surfaces and atmospheres of these objects to be determined and hints of these formation and evolutionary processes to be gleaned.

The Spitzer Space Telescope (SST) is, along with Hubble, a part of NASA's Great Observatories Program. Launched in 2003, SST carries an infrared array camera (IRAC), an infrared spectrograph (IRS), and a multiband imaging photometer. The IRS was split over four submodules with operational wavelengths of 5.3 to 40 $\mu m$[4] but has not been operational since Spitzer's helium coolant was depleted in 2009. Since the cool phase of Spitzer's mission ended, only the IRAC has remained operational, though with reduced capabilities. The Hubble WFC3 camera is currently delivering spectroscopic data at wavelengths shorter than 1.7 $\mu m$. Thus, at the time of writing this paper no space telescope capable of infrared spectroscopy beyond 1.7 $\mu m$ is operational.

The James Webb Space Telescope (JWST) is expected to be launched in March 2021. A near-infrared spectrometer (NIRSpec) and NIR camera are included within the instrument suite[5] and thus will provide the infrared capability that is currently missing (0.6 to 5.3 and 0.6 to 5.0 $\mu m$, respectively). In addition, the midinfrared instrument covers the wavelength range of 5 to 28 $\mu m$ and is capable of medium resolution spectroscopy.[6] However, a primary issue will be oversubscription, and not all the science cases will necessarily need the sensitivity and accuracy of JWST. Hence, whereas many opportunities exist for Solar System science with this mission (e.g., Refs. 7–9), a small space telescope, such as Twinkle, would offer an alternative for sources which are too bright to justify the use of JWST.

*Address all correspondence to Billy Edwards, E-mail: billy.edwards.16@ucl.ac.uk

We consider the possibility of using a small satellite in a low Earth orbit to perform spectroscopic remote-sensing observations of major Solar System objects. For the analysis, the instrument characteristics and performances of the Twinkle satellite[10–13] are adopted. Here, we study the feasibility of viewing major Solar System objects with Twinkle. First, the timescales over which major objects, such as planets, dwarf planets, and major asteroids, will be viewable is assessed. This analysis is then extended to planetary moons which potentially incur an additional observational constraint.

Having established a potential schedule for observing targets, the ability of Twinkle to obtain scientific data is evaluated. The sensitivity and saturation limits of each of the Twinkle's spectrometers is calculated for a given signal-to-noise ratio (SNR) and compared to the photon flux from a selection of Solar System bodies. The effect of combining multiple observations is also explored.

## 2 Twinkle

The Twinkle Space Mission is a new, fast-track satellite designed for launch in early 2022. It has been conceived for providing faster access to spectroscopic data from exoplanet atmospheres and Solar System bodies, but it is also capable of providing spectra of bright brown dwarfs and stars. Twinkle is equipped with a visible (0.4 to 1 $\mu$m) and infrared (1.3 to 4.5 $\mu$m) spectrometer (split into two channels at 2.42 $\mu$m). Twinkle has been designed with a telescope aperture of 0.45 m and will operate in a low Earth, Sun-synchronous orbit.[10,11]

Twinkle is a general observatory that is being managed by Blue Skies Space Ltd. Scientists will be able to purchase telescope time and Twinkle will provide on-demand observations of a wide variety of targets within wavelength ranges that are currently not accessible using other space telescopes or that are accessible only to oversubscribed observatories in the short-term future. Whereas it has been shown that Twinkle has significant capability for characterizing exoplanets,[13] the photometric and spectroscopic accuracies will also be well suited for observing a host of other targets including Solar System objects.

## 2.1 Payload

Twinkle is currently entering a phase B design review and thus the technical specifications stated here may change. Twinkle's scientific payload consists of a telescope with a 0.45-m aperture, a fine guidance sensor (FGS) and both a visible spectrometer and a NIRSpec, which can be operated simultaneously. The exoplanet light visible spectrometer is a visible spectrometer channel which is based on the UV and visible spectrometer (UVIS) flown on the ExoMars trace gas orbiter. For the Mars application, the UVIS instrument uses a dual telescope configuration: nadir (downward viewing of the surface for total atmospheric column measurements) and solar occultation observations (looking at the Sun through the atmosphere from orbit to measure vertical profiles). The telescopes are connected to a single spectrometer via a fiber-optic selector link. This telescope and selector system is not required in the Twinkle application as the spectrometer is positioned in the visible beam of the main Twinkle telescope.

The main modification to the spectrometer design is the use of an alternative grating and associated coatings to optimize the spectral range to the visible-to-NIR range between 0.4 and 1 $\mu$m with a resolving power of $R \sim 250$.[10] Other planned changes include a minor electronics component change on the detector board and relocation of the main electronics board stack to improve the thermal isolation and allow the detector to run at a lower temperature. Changes to the firmware code within the electronics will optimize the operations (e.g., CCD readout modes) and integration times for the Twinkle application.[10] This instrument is referred to as Channel 0 (Ch0). Several point spread functions (PSFs) across the visible channel are shown in Fig. 1. For the phase A study, an e2v CCD-230-42 detector is assumed for the visible channel, but this is currently under further discussion.

The design of Twinkle's NIRSpec is detailed in Wells.[12] As shown in Fig. 2, the NIRSpec will split the light into two channels (1.3 to 2.42 $\mu$m and 2.42 to 4.5 $\mu$m) to provide broadband coverage, while also ensuring appreciable spectral resolution. For shorter wavelengths ($\lambda < 2.42 \ \mu$m), the NIRSpec will have a resolving power of 250, whereas for longer wavelengths ($\lambda > 2.42 \ \mu$m), this will be reduced to 60. These channels are referred to as channels 1 and 2 (Ch1 and Ch2), respectively, and the spectrometer delivers a diffraction-limited image over both channels with the PSFs, as shown in Fig. 3. In the instrument design, a set of coupling lenslets is adopted to create an image of the aperture on the detectors. These lenses produce several spectra on the detector, with the spectrum from the star slit in the center with three spectra from the background slits on either side.[12] The two channels use different halves of the same detector (assumed to be produced by Selex in the phase A study). The format of the spectra on the detector is shown in Fig. 4. Owing to this layout, the two IR channels (Ch1 and Ch2) must be read out simultaneously, whereas the visible instrument (Ch0) can be read out independently.

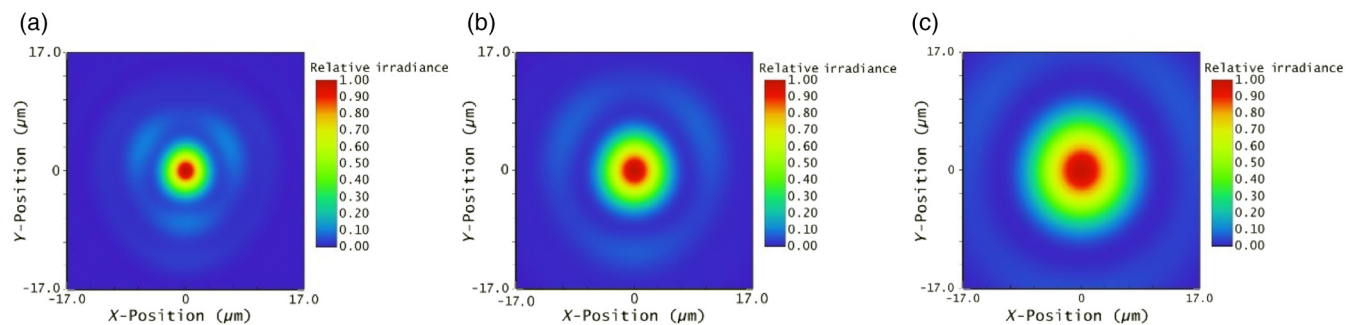

**Fig. 1** PSFs for the visible channel at (a) 0.4, (b) 0.6, and (c) 0.9 $\mu$m.

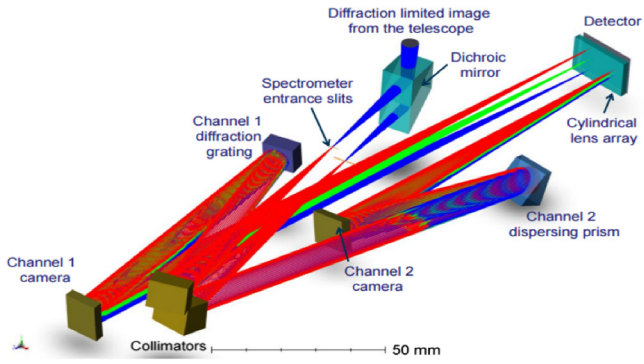

**Fig. 2** Perspective view of the IR instrument with the "entrance" prism seen in the top center of the picture. The rays enter vertically from the top and the two outputs (reflected and transmitted in the prism) exit the prism toward the bottom left of the viewer. The multitude of rays seen encompass three wavelengths per channel and the centers of the three fields (star + background fields on both sides of the star). Replication of Fig. 2-1 in Wells.[12]

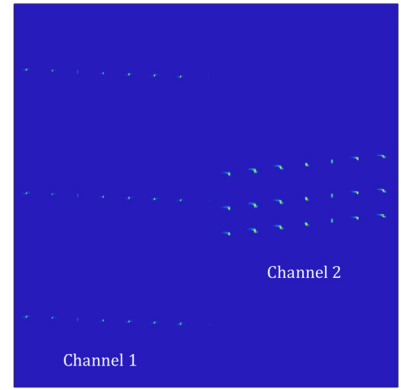

**Fig. 4** Format of spectra on the detector of the NIR instrument. In each channel, three spectra are produced from each background slit (S1–3 and S5–7) and a single spectrum from the star slit (S4) (located in the middle of the seven). $\lambda$-min and $\lambda$-max refer to the locations of the minimum and maximum wavelengths in each channel and the band center is situated at mid-$\lambda$. Adapted from Fig. 3-6 in Wells.[12]

The light incident on Twinkle's mirror will be directed through three entrance slits before being focused onto the instrumentation. Although the specifications have not been definitively established, Fig. 5 shows the expected angular sizes of Twinkle's slits for each spectral band, and these values were used for the following analysis.

The platform has a pointing accuracy of 1 arc min;[11] therefore, a FGS camera is to be used aboard Twinkle to facilitate precise pointing. The current design has a read-out frequency of 1 Hz and the FGS detector has a field of view of 6 × 6 arc min. Tip-tilt mirror control electronics will be utilized to keep the target within the slit for the duration of an observation and the pointing precision is expected to be in the order of 100 milliarcseconds (mas). A beam splitter is used to divide light between the visible spectrometer and the FGS. Further information on Twinkle, including publications describing the instrumentation, is available on the Twinkle website.[14]

## 2.2 Orbital Constraints and Target Visibility

The satellite will be placed in a low Earth (600 to 700 km), Sunsynchronous (dawn-dusk) polar orbit with a period of 90 to 100 min. The orientation of the satellite's orbit is constant with respect to the Sun but dictates that Twinkle's instrumentation must often be retargeted during an orbit to avoid Earth's limb. The boresight of the telescope will be pointed within a cone with a radius of 40 deg, which is centered on the anti-Sun vector (i.e., the ecliptic). Twinkle's instantaneous field of regard is complex to model but can be estimated as a half cone, which places a restriction on the targets which can be viewed at a given time. The field of regard could potentially be expanded to ±60 deg from the ecliptic for nondemanding targets.

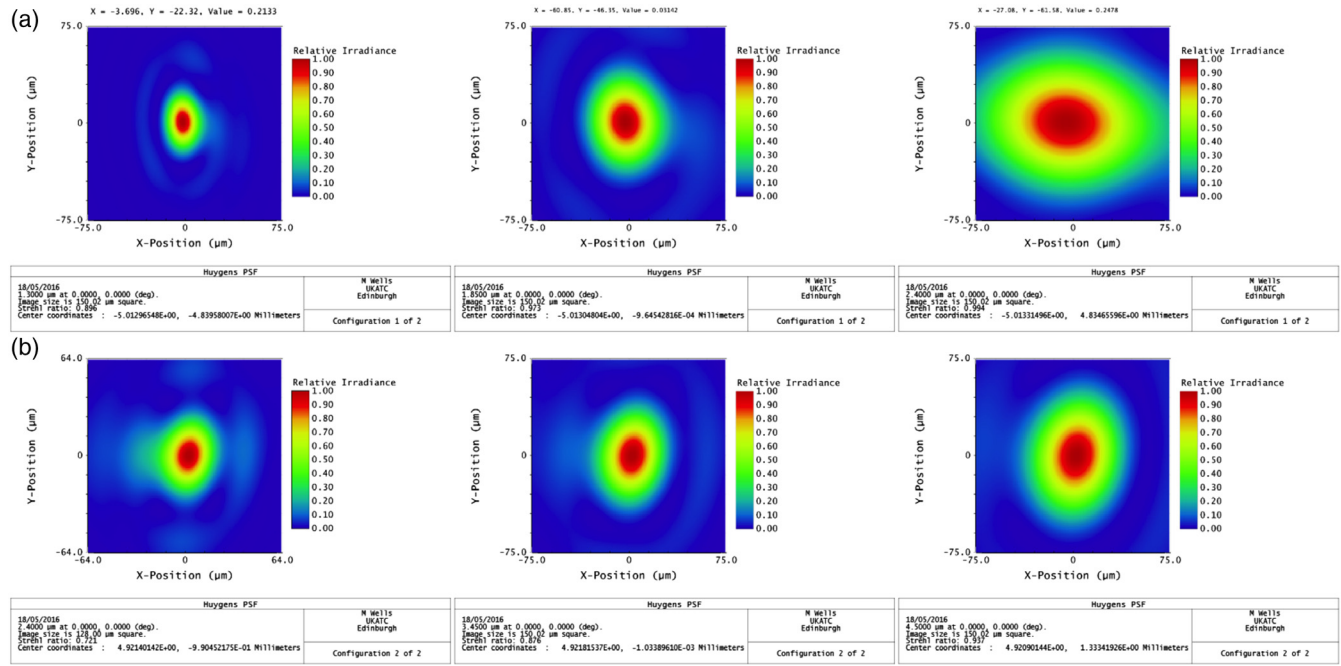

**Fig. 3** PSFs for channel 1 (a) at 1.3, 1.85, and 2.4 $\mu$m and channel 2 (b) at 2.4, 3.45, and 4.5 $\mu$m. The box is 10 × 10 pixels for all the PSFs. Replication of Fig. 3-3 in Wells.[12]

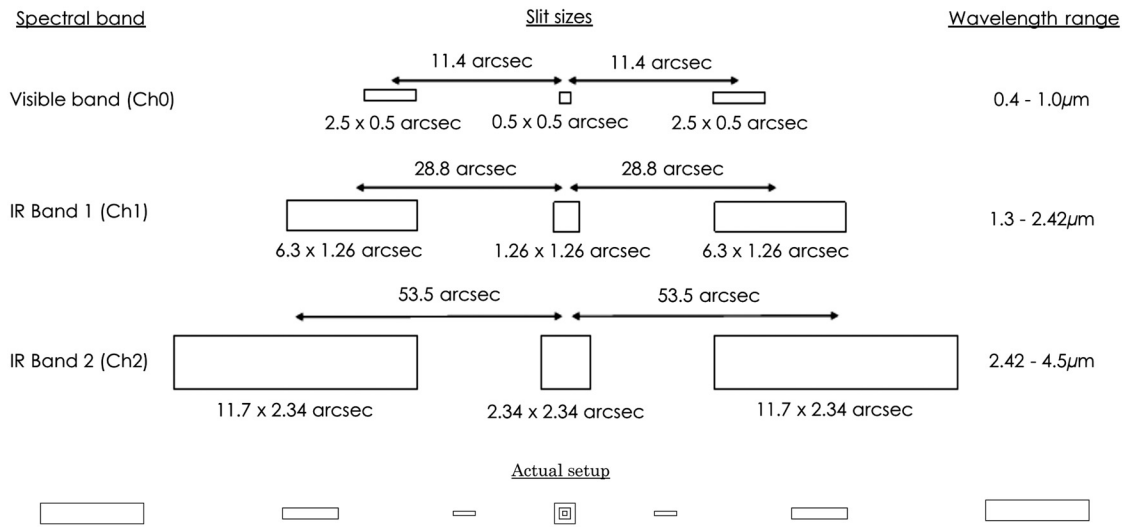

**Fig. 5** Top: Angular sizes of Twinkle's star and background slits (to scale). Bottom: The setup and actual separation between them shown to scale. As the star slits are overlapping, all three channels can observe the same target simultaneously.

The maximum amount of time a target outside of our Solar System could be observed during 1 year is around 81 days. The period of observation for objects within our Solar System could be constrained even further due to their proximity. The number of targets that Twinkle can view also depends on the agility of the satellite, i.e., its ability to alter its attitude to point at different targets and the time taken to achieve accurate pointing following this slew.

## 3 Target Availability

Twinkle has a design life of 7 years but, with no expendables, has the potential to operate for far longer. A precise launch date for the mission is still under discussion: for the purpose of this work, a "first light" date of January 1, 2022, was chosen and the following analysis was completed with a mission end date of January 1, 2032. Owing to the periodicity of observation windows for Solar System objects, the launch date has little effect on the availability of targets.

As stated, Twinkle's field of regard is centered on the anti-Sun vector and the time period during which targets can be viewed is limited. When considering the observations of celestial bodies, it is therefore key to determine how Twinkle's field of regard varies over time and when, and for how long, targets will be within this field.

Figure 6 shows the variation in the declination of the center of Twinkle's field of regard with right ascension. Note that, due to its Sun-synchronous orbit, the right ascension of the center of Twinkle's field of regard varies by 360 deg over the period of a year and that the declination variation is sinusoidal over the same period. The white circles indicate the extent of the telescope's field of regard at a given point. The center of the field of regard is located on the ecliptic at all times.

A model has been created to calculate Twinkle's field of regard for any date. Therefore, if the celestial coordinates of an object for a given date are known, it can be deduced whether this lies within Twinkle's observable range. The future ephemerides of the planets and other celestial bodies can be predicted with high accuracy by the Jet Propulsion Laboratory (JPL)'s Horizons system.[15] By comparing these predicted values to those of Twinkle over the lifetime of the mission, it can be determined when each object will be within Twinkle's field of regard and for how long.

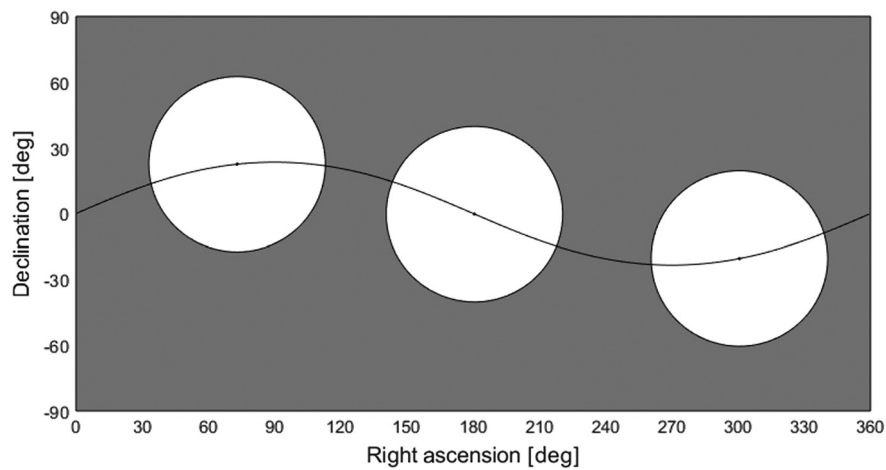

**Fig. 6** Variation in right ascension and declination of the center of Twinkle's field of regard and indication of its extent at a given time.

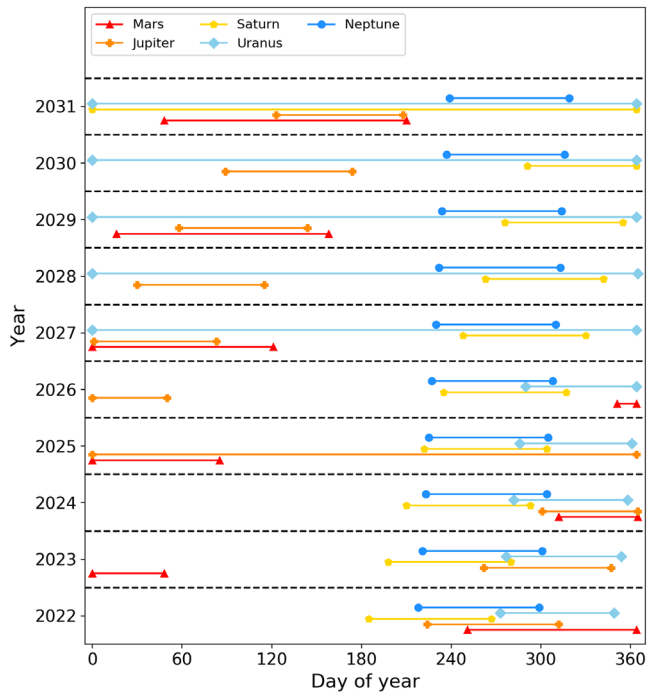

**Fig. 7** Periods for which the outer planets lie within Twinkle's field of regard from January 1, 2022, to January 1, 2032. Owing to Mars's close proximity to Earth and Twinkle's field of regard, Mars is only observable every other year.

### 3.1 Outer Planets

Figure 7 displays the observational periods for the outer planets over the period considered and shows that, generally, the further the planet is from Earth, the longer is the time period for which it can be observed. The outer planets are also found to have regular observation windows. As Twinkle's field of regard is centered on the anti-Sun vector it will not be possible to observe the inner planets.

### 3.2 Moons

Observations of one set of potential targets within the Solar System, i.e., planetary moons, incur an additional constraint. The moon of interest could be obscured behind the planet or be transiting across the face of the planet. In the case of the moon being behind the planet, obviously no spectral data can be obtained. When the moon is transiting across the planet, observations are likely to be subject to contamination by emissions from the planet. Therefore, for two segments of its orbit, a moon cannot be observed by Twinkle. The percentage of time for which a moon is viewable can be determined by calculating the time spent in front of, or behind, the host planet.

Figure 8 displays the orbit of Phobos and Triton over a period during which Mars and Neptune lie within Twinkle's field of regard. It should be noted that difficulties may be encountered if observations are performed when the moon is close to the limb of its host planet due to stray light. As the angular size of planetary moons is generally small compared to the size of Twinkle's slits, a gap between a moon and the planet's limb is desirable to reduce this stray light. The appropriate size of this gap is likely to vary depending on the target. Increasing the gap reduces the likelihood of the planet contaminating the spectra. However, requiring a large gap reduces the time for which a moon can be viewed.

If one wishes for separation between the edge of the slit and the planet, a gap of 2.5 arc sec could be considered minimum as (assuming the slit is centered on the moon) this provides a gap of

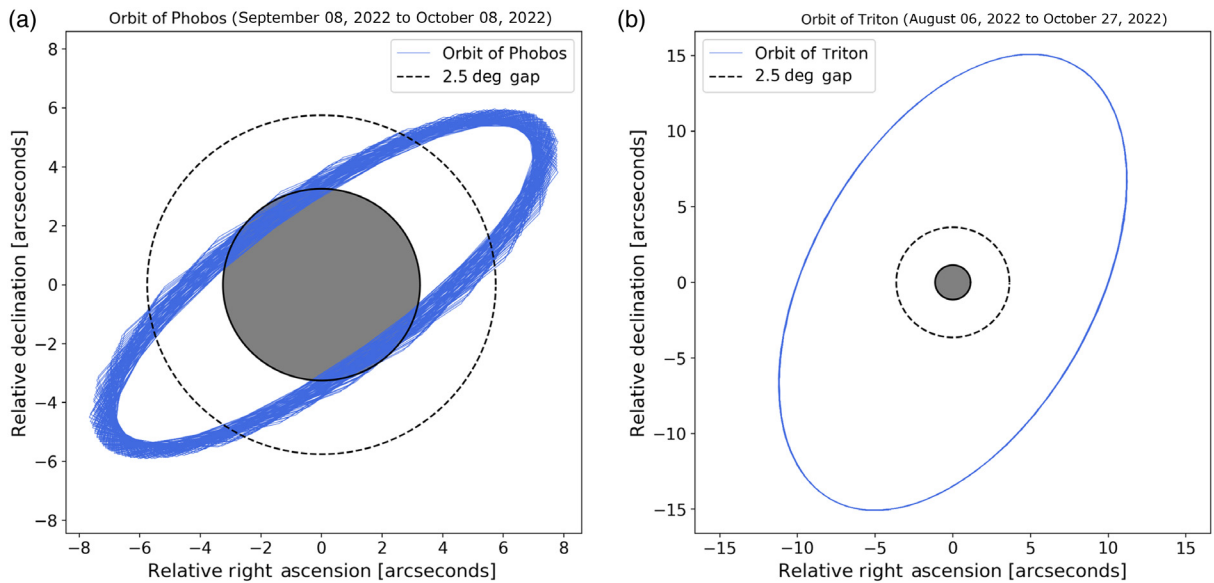

**Fig. 8** (a) Orbit of Phobos during some of the period for which Mars is within Twinkle's field of regard. (b) Orbit of Triton when Neptune is potentially viewable. The host planets are represented by the shaded central circles (to scale). Note that the orbital period of Phobos and Triton are 7.38 h and 5.88 days, respectively.

at least 1 arc sec between the edge of the slit and the planetary limb.

It is found that, in the majority of cases, the moons are viewable for large proportions of their orbits but that observable periods change over longer time frames due to the changing inclination of the moon's orbit with respect to an observer on Earth.

### 3.3 Dwarf Planets and other Major Celestial Bodies

In the same way as for planets, the analysis was conducted for other celestial bodies. The possibility of viewing dwarf planets as well as major asteroids and Trojans was explored and the observation periods for these objects are shown in Figs. 14 and 15. We find that the more distant a target, the smaller is the variation in observable period over the lifetime of the mission.

## 4 Instrumentation Performance and Data Quality

Once it has been concluded that an object is viewable, the performance of Twinkle's instrumentation and the potential quality of scientific data must be ascertained. The specifications of the instrumentation have not been definitively established and any changes will impact the conclusions drawn here. However, the following analysis is readily adaptable to new instrumentation parameters.

### 4.1 Angular Size

As discussed, the slits for each spectral band have different angular sizes. From JPL's Horizons system, the angular diameter of a target at a given time can be found. As demonstrated, the observation windows for viewing a target can be determined and thus the average angular diameter of a target when viewable can be calculated. This angular size can be compared to the angular size of Twinkle's star and background slits for each spectral band to ascertain whether the target can be viewed in its entirety in one observation.

If a target is too large to view in one observation, an estimate of the number of observations required to fully map the target is critical to both the total time needed to observe the entire target and the average SNR associated with each observation. In the case where only one observation is needed, all photons from the target's visible disc are collected in one spectrum. If multiple spectra are taken, then the number of photons received from a given observation area is dependent on the total number of photons from the target and the number of observations needed to map the target (i.e., if four observations are needed, a constant surface brightness is assumed and thus that a quarter of the total number of photons contributes to each spectrum).

An estimate to quantify the required number of observations is given by dividing the target's angular area by that of the viewing slit of Twinkle. If the planet were to be totally mapped, this lower bound is unachievable as the shape of the viewing slit does not tessellate to form a circle. However, for the purpose of defining the fraction of photons from the viewed segment of target, this approximation is valid. Here, we assume that the star slit is used for observations, and the number required to map the entire visible face of various targets is shown in Table 1.

The photon flux received may vary due to planetary features (e.g., the bands on Jupiter) and may also be affected by limb darkening. Hence, the exposure time needed for a desired

**Table 1** Approximate number of observations required for each instrument to cover the entire visible face of a target using the star slit.

| Target | Angular diameter (arc sec) | Approximate number of observations | | |
|---|---|---|---|---|
| | | Ch0 | Ch1 | Ch2 |
| Mars | 16.02 | 806 | 127 | 37 |
| Jupiter | 46.17 | 6697 | 1055 | 306 |
| Europa | 1.03 | 4 | 1 | 1 |
| Saturn | 16.74 | 881 | 139 | 41 |
| Titan | 0.74 | 2 | 1 | 1 |
| Uranus | 3.78 | 45 | 8 | 1 |
| Neptune | 2.35 | 18 | 3 | 1 |
| Triton | 0.13 | 1 | 1 | 1 |
| Pluto | 0.10 | 1 | 1 | 1 |

SNR could fluctuate depending on the type of observation carried out. However, this difference should be relatively small due to the brightness of such a target.

The targets in question are also rotating, the rate of which will dictate the length of time necessary to allow Twinkle to gain spectral data for the whole surface of the object. In general, these rotation times are low compared to the observation window, allowing plenty of time for observations of any side of the object.

### 4.2 Estimate of the Signal Received from a Target

Assuming that the systematic noise characteristics of the instrument are constant for any target, the exposure time needed for a desired SNR is dictated by the number of photons received per second. When observing a Solar System target, the flux received originates from two sources, the reflection of solar radiation and the radiation emitted from the target itself. Thus, the flux received at Earth from a target body is given as

$$F_{\text{Target}} = F_{\text{Reflected}} + F_{\text{Emitted}}. \quad (1)$$

The flux from the target due to reflected solar radiation is given as

$$F_{\text{Reflected}} = \frac{\text{Solar Flux at Target} \, (\text{W m}^{-2}\,\text{nm}^{-1}) \times p_v \times \pi R_T^2}{2\pi D_{E-T}^2}, \quad (2)$$

where $R_T$ is the radius of the body, $D_{E-T}$ is the separation between the target and Earth, and the $p_v$ is the albedo, the fraction of incident radiation, which is reflected in the direction of the observer. It is assumed that the visible face of the body reflects solar radiation uniformly over a half sphere of radius $D_{E-T}$.

The previously described analysis determined when a target could be viewed by Twinkle and was used to calculate the average Sun-target separation as well as the Earth-target distance.

We used the ASTM E-490 solar spectral irradiance[16] to calculate the reflected light from the target body.

The geometric albedo of the target, as well as its radius, can be acquired from a variety of sources and thus the photon flux of solar radiation reflected by the target at the observer can be obtained. The values used here are depicted in Table 3 in Sec. 7. The amount of radiation reflected by a target in the direction of the observer depends on the phase angle, the angle formed between the Sun, the target, and the observer. The geometric albedo, $q$, is used to calculate the albedo of a target from the following equation:

$$A = p_v \times q(\chi),  \qquad (3)$$

where $q(\chi)$ is the phase integral which, for planetary bodies, can be estimated as that for a diffuse sphere and is given as

$$q(\chi) = \frac{2}{3}\left[\left(1 - \frac{\chi}{\pi}\right)\cos(\chi) + \frac{1}{\pi}\sin(\chi)\right],  \qquad (4)$$

where $\chi$ is the phase angle.[41] Given that Twinkle's field of regard is centered on the ecliptic and that the orbits of major bodies have small inclinations, observations will generally occur at low phase angles and thus a phase angle of 10 deg has been assumed (i.e., $q \approx 0.657$). Geometry effects such as coherent backscatter have been omitted.

Modeling the target as a black body, the flux received at Earth due to emission from the target can be determined. Here, the effective temperature has not been derived but instead literature values have been used. For a solid body, the surface temperature is used, and for the gaseous planets the temperature at 1 bar is taken. It is assumed that the body radiates uniformly over a sphere with a radius equal to the average Twinkle-target distance.

For all targets in the shorter infrared band (Ch1) and the visual band (Ch0), as well as the vast majority of targets in the longer infrared band (Ch2), it is found that the contribution of emitted radiation is negligible. The exception is the Martian satellites, which have an emission contribution of ~18% in the second infrared band.

### 4.3 Instrumental Performance and Noise

Not all incident photons will be detected due to instrument inefficiencies. The optical and quantum efficiencies are shown in Table 2 along with the instrument plate scale.

A model was created to estimate the noise contributions from various sources. The detector is assumed to be cooled to 70 K,

**Table 2** Instrument properties over each channel.

| Instrument | Ch0 | Ch1 | Ch2 |
|---|---|---|---|
| Property | (0.4 to 1 μm) | (1.3 to 2.42 μm) | (2.42 to 4.5 μm) |
| Optical efficiency | 0.80 | 0.45 | 0.61 |
| Quantum efficiency | 0.7 | 0.7 | 0.7 |
| Plate scale (arc sec /μm) | 0.007 | 0.027 | 0.7 |

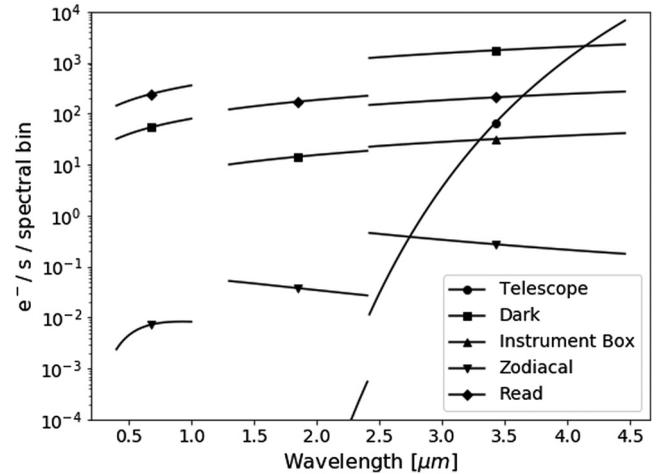

**Fig. 9** Number of electrons per second per-spectral bin from different noise sources when operating at $R \sim 250$ ($\lambda < 2.42$ μm) and $R \sim 60$ ($\lambda > 2.42$ μm).

whereas the telescope has been modeled at 180 K. Excluding the target signal, Fig. 9 shows the noise levels from each source with the dark current dominating most wavelengths for long exposures, although the telescope is dominant at longer wavelengths. For short exposures, the read noise dominates and the contribution from the instrument box is only significant in Ch2 ($\ll 10^{-4}$ for the other channels). Owing to this systematic noise, the detector will saturate at longer wavelengths within ~600 s, even for faint targets.

### 4.4 Determining Observability

By setting a requirement of SNR = 10 (or 100), we calculate the ability of Twinkle to observe an object at the highest resolving power for each channel using the star slit. The photon flux (photons/m²/s) received from a target is calculated across each spectral bin. Within each channel, the photon flux varies due to the variance in solar output with wavelength. Thus, for ease of representation, the average flux per spectral bin is calculated for each channel.

As discussed, the detector noise levels have been estimated on a per-spectral bin basis. Therefore, the noise contributing in a given spectral bin is used to obtain the minimum photon flux needed to meet the SNR requirement within a given integration time by rearranging the SNR equation:

$$SNR = \frac{N * QE * \eta}{\sqrt{N_\gamma + (N * QE * \eta)}}\sqrt{t_{EXP}},  \qquad (5)$$

where $t_{EXP}$ is the exposure time, $N_\gamma$ is the total systematic noise (per-spectral bin per second and calculated from the noise sources in Fig. 9), $QE$ is the quantum efficiency, $\eta$ is the optical efficiency, and $N$ is the number of electrons (per-spectral bin per second) from the target.

By comparing the photon flux from a target with the minimum required to achieve SNR = 10 (or 100), the Solar System objects which could be observed by Twinkle are determined. For some bright sources, saturation can be an issue and thus the maximum photon flux from a target that could be observed for a given integration time is also calculated. We assume a maximum continuous integration time of 300 s.

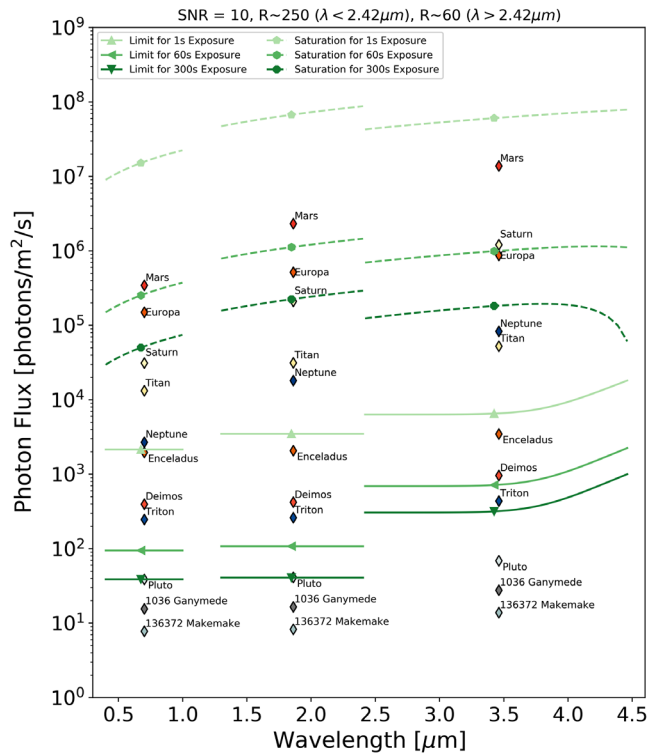

**Fig. 10** The average photon flux received per-spectral bin at Earth for various Solar System bodies at their average distance during observable periods with Twinkle. In addition, the sensitivity and saturation limits of Twinkle are plotted for single observations with various exposure times using the star slit assuming that a SNR > 10 is required. The corresponding figure for SNR = 100 is provided in Sec. 7.

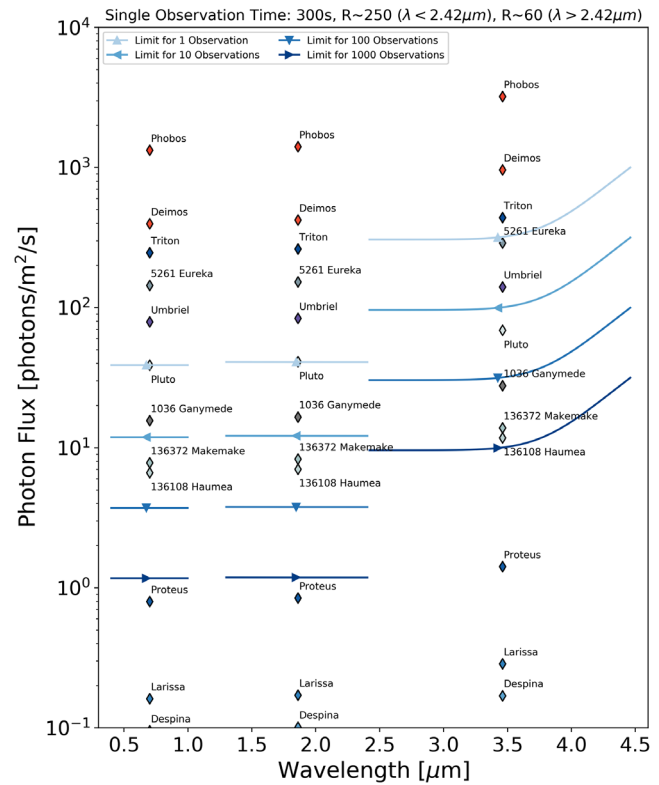

**Fig. 11** The average photon flux received per-spectral bin at Earth for various Solar System bodies at their average distance during observable periods with Twinkle. In addition, the sensitivity limits of Twinkle are plotted for a given number of 300 s observations with the star slit.

These sensitivity and saturation limits are plotted in Fig. 10 and, if an object lies between these limits for a given exposure time, Twinkle can achieve spectra at the instrumentation's native resolution with an SNR > 10. We find that very bright objects such as Mars could only be observed with short exposure times due to saturation of the detectors. Targets such as Neptune could be observed in around 1 s, whereas spectra of dimmer objects (e.g., Deimos) could be obtained in 60 s. A full table of results is shown in Sec. 7 as well as the sensitivity limits for SNR = 100 in Sec. 8 (Fig. 13).

By combining multiple observations, the faintest object that could be observed by Twinkle can be increased. We find that with <10 observations with exposure times of 300 s, Pluto could be observed at Twinkle's native resolution, as shown in Fig. 11. The sensitivity limit of Twinkle could be further increased by binning down the spectra, reducing the resolution but increasing the number of photons per-spectral bin.

Thus, we find that many celestial bodies could be observed at a high spectral resolution ($R \sim 250$, $\lambda < 2.42 \ \mu m$; $R \sim 60$, $\lambda > 2.42 \ \mu m$) with short exposure times (≪300 s). On the other hand, some bodies are too small, faint, or distant and are found to require many observations. The potential exists to bin the spectra to lower resolutions to increase the SNR and thus allow faint objects to be observed.

# 5 Discussion

This first iteration of assessing Twinkle's performance for Solar System science has shown that many objects are potentially observable with Twinkle. Twinkle is currently entering a phase B design review and thus the technical specifications may change. An updated analysis will be published when the design is finalized.

## 5.1 Suitability of Targets

From the analysis presented here, it can be concluded that, for each object, an ideal viewing period will exist and this is when the target is closest to the Sun (and Earth), at a low phase angle, close to the center of Twinkle's field of regard. This will ensure the highest photon flux. The photon fluxes calculated here are averages of the expected performance. Thus, an observation at a specified time may achieve better or worse performance than predicted here. The capability of Twinkle to view a target also varies over the spectral bins. Targets have been assessed by the number of observations required to achieve high-resolution spectroscopic data with a SNR of 10. If this level of resolution or quality is not needed, the capability of Twinkle to observe this object may change. For example, in the infrared, many spectral features can be broad and thus a resolving power of 250 (or 60) may not be necessary.

It is, therefore, concluded that, although some targets will not be suitable for observations with Twinkle, there is the potential to observe, with a relatively small space telescope, a significant number of objects. The resolution and quality of data achievable will vary with each spectral bin and with the length and number of observations undertaken. Reducing the required SNR will considerably increase the ease for which Twinkle could obtain data from a target.

For planets and large moons, the integration times are, in many cases, shown to be short enough that spectroscopic observations of these bodies can be undertaken in a single exposure. For faint, small, or distant targets, the integration times are longer and in several cases are too large to obtain high-quality data in one observation. Combining multiple observations of a target will increase the data quality obtainable as will binning the spectra to a lower resolution.

We find that Solar System objects generally have long observing windows (up to 80 days) each year and these potential observing periods are also periodic over the timescale of the mission. As, in general, the observation window for an object is greater than the rotation period of the target, Twinkle could obtain global views of spatial variations. By reobserving an object over an extended period of time (i.e., years), observations with Twinkle could also be used to search for temporal variations.

## 5.2 Tracking Capability of the Fine Guidance Sensor

The detailed tracking performance of the FGS will ultimately depend on platform pointing accuracy. However, it is expected that the wide field of view of the FGS camera will allow bright sidereal targets to be tracked. Current simulations suggest that this will be possible for targets with visible magnitude of 15 or brighter. Further investigation is needed to fully ascertain the capability of the FGS, and this will be performed as part of the phase B study. For fainter targets, tracking could be simulated by scanning linear track segments. These linear track segments are linear in equatorial coordinate space; they are commanded as a vector rate in J2000 coordinates, passing through a specified RA and Dec at a specified time. The coordinates of the target can be obtained from services, such as JPL's Horizons system. This method of tracking is by no means simple but has been employed on Spitzer (and will be for JWST). Including such a capability would be nontrivial but, given the current status of the mission, there is time to include and refine this capacity.

The maximum tracking rate of Twinkle is also subject to further investigation. During the ExploreNEO program, Spitzer achieved a maximum tracking rate of 543 mas/s[42] and JWST will be capable of 30 mas/s.[7] The rate of Mars is around 30 mas/s with the outer planets having smaller rates of motion. Twinkle's FGS will be capable of tracking objects of this speed and the current design suggests that much higher rates could be tolerated. Therefore, for major Solar System bodies, the rate of motion should not be prohibitive. Overall, the scope of this paper is to understand the capability of Twinkle's spectrometers to observe major Solar System bodies. By showing that Twinkle can obtain spectra of a wide variety of interesting targets, a requirement is subsequently placed on the FGS to be able to track these objects to allow for this science case to be implemented.

## 5.3 Background Slits

Here, the capabilities of the star slit to observe Solar System objects have been assessed and, in the normal operation mode, the background slits are used to provide measurements of zodiacal light in the direction of observation which is then subtracted from the observation of the target. When observing a planet, the zodiacal component will be small compared to the planetary contribution.

Therefore, the background slits could instead be used to provide spectra of a strip of the target. This increases the area covered in one observation and decreases the exposure time required, although spatial resolution is also reduced. The star slit (or other background slit) could potentially be used to subtract any background. Utilizing the background slit in this way would rapidly decrease the time taken to fully observe the visible face of a large target.

There may, however, be issues when using Twinkle in this way. The telescope is designed such that the FGS is used for fine pointing. In normal operations, this will fix the star on pixels of the FGS CCD and keep its location fixed. As the spacecraft moves, due to its orbit and other perturbations, the FGS mirror tilts to keep the target positioned on this pixel. This ensures that the slit is pointed at the target for the entirety of the observation. For small targets, such as moons or asteroids, for which only one observation is required to map, the methodology is the same.

However, if multiple spectra need to be taken to map the object then this would require Twinkle to not only keep the telescope pointed at a given location but also be able to fine point within this object. This is vastly different from the primary mode of operation and thus a different mode would need to be devised. The slew time between these observations will also need to be accounted for in the mapping time and the precision to which these adjustments can be accomplished must be determined. Though potentially problematic, this mapping technique is certainly not inconceivable.

The background slits could also mitigate for the contribution of the planet when observing moons. The background slit could be used to observe the moon with its orientation being perpendicular to the planet-satellite direction, as shown in Fig. 12, with the central background spectra focused on the moon. For moons, and other small objects, care will have to be taken to ensure that other background objects, such as bright stars, do not lie within the field of view, contributing to the received spectra. In addition, as all three slits image onto the same detector, the other slits must be positioned away from very bright objects, such as the host planet of a moon, to avoid saturation. A final consideration is that the background slits observe different regions of the sky so multiple observations would be needed to obtain full spectral coverage.

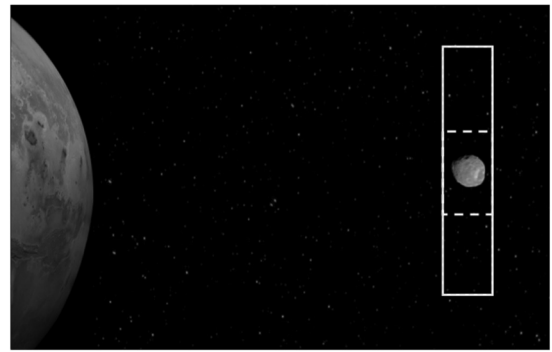

**Fig. 12** Potential use of background slit to view moons and mitigate for stray light from the host planet (not to scale). Original images credit: NASA.

### 5.4 Potential Impact

Twinkle, therefore, offers a capability that is not currently available: space-based spectroscopic observations of Solar System objects in the infrared beyond 1.7 $\mu$m. Unlike ground-based telescopes, Twinkle will not be inhibited by Earth's atmosphere and thus will be capable of detecting water features in the infrared that would be extremely difficult from the ground. In addition to hydration features, many silicates and organics have absorption features within Twinkle's 0.4- to 4.5-$\mu$m range. These are key for understanding the dispersion of water, metals, and organics throughout the Solar System and provide insights into planetary formation and evolution. Several $CO_2$ ice bands also lie within Twinkle's wavelength coverage (e.g., those around 2 $\mu$m and the absorption feature at 4 to 4.4 $\mu$m, which is blocked by telluric $CO_2$), which have been detected in various bodies including the moons of Jupiter,[43,44] Saturn,[45] and Uranus.[46] Additional molecules found in these satellites that may be detectable with Twinkle include condensed $O_2$[47] and $SO_2$.[48] Combining spectroscopic observations of major bodies with small bodies (asteroids and comets) enables a detailed view of the primitive and current state of the Solar System. Thus, further work will seek to understand the capability of Twinkle to observe small bodies within the Solar System.

### 6 Conclusions

The capability of the Twinkle space telescope to observe Solar System objects in visible and infrared wavelengths has been assessed.

By comparing the celestial coordinates of potential targets with Twinkle's field of regard it has been found that, although the orbital characteristics of Twinkle impose constraints, the potential observation windows are still large. The duration and frequency of these observable periods vary, as does the trajectory of the object across Twinkle's field of regard. Planetary moons incur an additional constraint but obscuration by their host planet is found to be minor in the vast majority of cases, although some observations may be hindered by stray light.

Solar System targets, for which Twinkle's capabilities allow for the acquisition of high-quality, high-resolution spectroscopic data within a single observation is found to incorporate planets and some larger moons. The potential also exists for observations of smaller moons and large asteroids at high resolution, whereas photometric observations should be possible for a vast number of objects. The capability of Twinkle to observe these objects varies with wavelength and, as the majority of photons received from bodies are reflected solar radiation, the longer infrared band has the lowest capabilities for viewing fainter objects, which is also due to higher instrumentation noise.

An obvious way of increasing the sensitivity limit is to combine multiple observations and this has been shown to expand the targets which Twinkle could observe. For each target an optimal approach\ is likely to exist and, by varying the resolving power, length or number of observations and the required SNR, the ability of Twinkle can be assessed on a case-by-case basis.

Twinkle is found to have significant potential for viewing Solar System objects. High-resolution spectroscopic data for brighter targets such as the planets and larger, nearer moons can be readily acquired, although saturation limits may have to be considered. In addition, lower-resolution spectroscopic observations of smaller, fainter, or more distant objects could be within the reach of Twinkle's capabilities.

### 7 Appendix A: Full Table of Results

Table 3 displays the target characteristics used here and the average flux across each of Twinkle's instrument channels. For comparison, Table 4 contains the sensitivity and saturation limits of Twinkle.

**Table 3** Average photon flux (photons/m$^2$/s) per-spectral bin for targets considered here and the assumed parameters used in the calculation. Many of the albedos have been acquired for the JPL Solar System dynamics service and the original sources have been cited wherever possible.

| Target name | Radius (m) | Geometric albedo | Temperature (K) | Flux (photons/m$^2$/s) | | |
| --- | --- | --- | --- | --- | --- | --- |
| | | | | Ch0 | Ch1 | Ch2 |
| Mars | $3.39 \times 10^{06}$ | 0.15[17] | 210 | $3.44 \times 10^{05}$ | $2.32 \times 10^{06}$ | $1.38 \times 10^{07}$ |
| Phobos | $1.11 \times 10^{04}$ | 0.071[18] | 233 | $1.33 \times 10^{03}$ | $1.41 \times 10^{03}$ | $3.21 \times 10^{03}$ |
| Deimos | $6.20 \times 10^{03}$ | 0.068[19] | 233 | $3.97 \times 10^{02}$ | $4.21 \times 10^{02}$ | $9.62 \times 10^{02}$ |
| Jupiter | $6.99 \times 10^{07}$ | 0.52 | 165 | $1.12 \times 10^{05}$ | $7.52 \times 10^{05}$ | $4.35 \times 10^{06}$ |
| Ganymede | $2.63 \times 10^{06}$ | 0.435[20] | 103 | $9.34 \times 10^{04}$ | $6.29 \times 10^{05}$ | $1.53 \times 10^{06}$ |
| Callisto | $2.41 \times 10^{06}$ | 0.18[20] | 118 | $3.86 \times 10^{04}$ | $2.60 \times 10^{05}$ | $5.32 \times 10^{05}$ |
| Io | $1.82 \times 10^{06}$ | 0.625[21] | 118 | $1.34 \times 10^{05}$ | $6.29 \times 10^{05}$ | $1.05 \times 10^{06}$ |
| Europa | $1.56 \times 10^{06}$ | 0.7[22] | 113 | $1.50 \times 10^{05}$ | $5.18 \times 10^{05}$ | $8.67 \times 10^{05}$ |
| Himalia | $8.50 \times 10^{04}$ | 0.675 | 124 | $1.41 \times 10^{03}$ | $1.49 \times 10^{03}$ | $2.50 \times 10^{03}$ |
| Amalthea | $8.35 \times 10^{04}$ | 0.091[23] | 160 | $1.82 \times 10^{02}$ | $1.93 \times 10^{02}$ | $3.24 \times 10^{02}$ |
| Thebe | $4.93 \times 10^{04}$ | 0.047[23] | 124 | $3.27 \times 10^{01}$ | $3.47 \times 10^{01}$ | $5.81 \times 10^{01}$ |

**Table 3** (*Continued*).

| Target name | Radius (m) | Geometric albedo | Temperature (K) | Flux (photons/m$^2$/s) | | |
|---|---|---|---|---|---|---|
| | | | | Ch0 | Ch1 | Ch2 |
| Elara | $4.30 \times 10^{04}$ | 0.035 | 124 | $1.84 \times 10^{01}$ | $1.95 \times 10^{01}$ | $3.27 \times 10^{01}$ |
| Saturn | $5.82 \times 10^{07}$ | 0.47 | 134 | $3.12 \times 10^{04}$ | $2.10 \times 10^{05}$ | $1.21 \times 10^{06}$ |
| Titan | $2.57 \times 10^{06}$ | $0.2^{24}$ | 94 | $1.33 \times 10^{04}$ | $3.14 \times 10^{04}$ | $5.26 \times 10^{04}$ |
| Rhea | $7.64 \times 10^{05}$ | $0.949^{25}$ | 76 | $1.24 \times 10^{04}$ | $1.32 \times 10^{04}$ | $2.21 \times 10^{04}$ |
| Iapetus | $7.36 \times 10^{05}$ | $0.6^{24}$ | 110 | $7.29 \times 10^{03}$ | $7.74 \times 10^{03}$ | $1.29 \times 10^{04}$ |
| Dione | $5.62 \times 10^{05}$ | $0.85^{25}$ | 87 | $6.02 \times 10^{03}$ | $6.39 \times 10^{03}$ | $1.07 \times 10^{04}$ |
| Tethys | $5.31 \times 10^{05}$ | $1.229^{25}$ | 86 | $7.77 \times 10^{03}$ | $8.24 \times 10^{03}$ | $1.38 \times 10^{04}$ |
| Enceladus | $2.52 \times 10^{05}$ | $1.375^{25}$ | 75 | $1.96 \times 10^{03}$ | $2.08 \times 10^{03}$ | $3.48 \times 10^{03}$ |
| Mimas | $1.98 \times 10^{05}$ | $0.962^{25}$ | 64 | $8.46 \times 10^{02}$ | $8.97 \times 10^{02}$ | $1.50 \times 10^{03}$ |
| Hyperion | $1.35 \times 10^{05}$ | $0.3^{24}$ | 93 | $1.23 \times 10^{02}$ | $1.30 \times 10^{02}$ | $2.18 \times 10^{02}$ |
| Uranus | $2.54 \times 10^{07}$ | $0.51^{26}$ | 76 | $7.96 \times 10^{03}$ | $5.36 \times 10^{04}$ | $3.09 \times 10^{05}$ |
| Titania | $7.89 \times 10^{05}$ | $0.27^{26}$ | 70 | $1.85 \times 10^{02}$ | $1.96 \times 10^{02}$ | $3.29 \times 10^{02}$ |
| Oberon | $7.61 \times 10^{05}$ | $0.23^{26}$ | 75 | $1.47 \times 10^{02}$ | $1.56 \times 10^{02}$ | $2.60 \times 10^{02}$ |
| Umbriel | $5.85 \times 10^{05}$ | $0.21^{26}$ | 75 | $7.91 \times 10^{01}$ | $8.40 \times 10^{01}$ | $1.41 \times 10^{02}$ |
| Ariel | $5.79 \times 10^{05}$ | $0.39^{26}$ | 60 | $1.44 \times 10^{02}$ | $1.53 \times 10^{02}$ | $2.56 \times 10^{02}$ |
| Miranda | $2.36 \times 10^{05}$ | $0.32^{26}$ | 60 | $1.96 \times 10^{01}$ | $2.08 \times 10^{01}$ | $3.49 \times 10^{01}$ |
| Neptune | $2.46 \times 10^{07}$ | 0.41 | 72 | $2.70 \times 10^{03}$ | $1.82 \times 10^{04}$ | $8.30 \times 10^{04}$ |
| Triton | $1.35 \times 10^{06}$ | $0.719^{27}$ | 58 | $2.47 \times 10^{02}$ | $2.62 \times 10^{02}$ | $4.38 \times 10^{02}$ |
| Proteus | $2.10 \times 10^{05}$ | $0.096^{28}$ | 51 | $7.97 \times 10^{-01}$ | $8.46 \times 10^{-01}$ | $1.42 \times 10^{00}$ |
| Nereid | $1.70 \times 10^{05}$ | $0.155^{29}$ | 51 | $8.44 \times 10^{-01}$ | $8.95 \times 10^{-01}$ | $1.50 \times 10^{00}$ |
| Larissa | $9.70 \times 10^{04}$ | $0.091^{28}$ | 51 | $1.61 \times 10^{-01}$ | $1.71 \times 10^{-01}$ | $2.86 \times 10^{-01}$ |
| Galatea | $8.80 \times 10^{04}$ | $0.079^{28}$ | 51 | $1.15 \times 10^{-01}$ | $1.22 \times 10^{-01}$ | $2.05 \times 10^{-01}$ |
| Despina | $7.50 \times 10^{04}$ | $0.09^{28}$ | 51 | $9.54 \times 10^{-02}$ | $1.01 \times 10^{-01}$ | $1.69 \times 10^{-01}$ |
| Thalassa | $4.10 \times 10^{04}$ | $0.091^{28}$ | 51 | $2.88 \times 10^{-02}$ | $3.06 \times 10^{-02}$ | $5.12 \times 10^{-02}$ |
| Naiad | $3.30 \times 10^{04}$ | $0.072^{28}$ | 51 | $1.48 \times 10^{-02}$ | $1.57 \times 10^{-02}$ | $2.62 \times 10^{-02}$ |
| 1 Ceres | $4.70 \times 10^{05}$ | $0.09^{30}$ | 168 | $3.88 \times 10^{01}$ | $4.12 \times 10^{01}$ | $6.89 \times 10^{01}$ |
| Pluto | $1.19 \times 10^{06}$ | 0.3 | 50 | $6.63 \times 10^{00}$ | $7.03 \times 10^{00}$ | $1.18 \times 10^{01}$ |
| 136108 Haumea | $5.75 \times 10^{05}$ | $0.84^{31}$ | 50 | $7.81 \times 10^{00}$ | $8.28 \times 10^{00}$ | $1.39 \times 10^{01}$ |
| 136372 Makemake | $7.15 \times 10^{05}$ | $0.81^{32}$ | 42 | $2.23 \times 10^{-02}$ | $2.37 \times 10^{-02}$ | $3.97 \times 10^{-02}$ |
| 136199 Eris | $1.16 \times 10^{05}$ | $0.96^{33}$ | 43 | $1.44 \times 10^{02}$ | $1.53 \times 10^{02}$ | $2.89 \times 10^{02}$ |
| 5261 Eureka | $1.30 \times 10^{03}$ | $0.39^{34}$ | 250 | $1.62 \times 10^{-03}$ | $1.72 \times 10^{-03}$ | $2.88 \times 10^{-03}$ |
| 433 Eros | $8.42 \times 10^{03}$ | $0.25^{35}$ | 230 | $3.03 \times 10^{05}$ | $3.21 \times 10^{05}$ | $5.38 \times 10^{05}$ |
| 4 Vesta | $2.63 \times 10^{05}$ | $0.38^{36}$ | 150 | $6.95 \times 10^{04}$ | $1.26 \times 10^{05}$ | $2.11 \times 10^{05}$ |
| 2 Pallas | $2.56 \times 10^{05}$ | $0.16^{37}$ | 164 | $6.45 \times 10^{04}$ | $6.84 \times 10^{04}$ | $1.15 \times 10^{05}$ |

**Table 3** (*Continued*).

| Target name | Radius (m) | Geometric albedo | Temperature (K) | Flux (photons/m²/s) | | |
|---|---|---|---|---|---|---|
| | | | | Ch0 | Ch1 | Ch2 |
| 10 Hygiea | $2.16 \times 10^{05}$ | $0.07^{38}$ | 164 | $9.24 \times 10^{03}$ | $9.80 \times 10^{03}$ | $1.65 \times 10^{04}$ |
| 1036 Ganymed | $1.58 \times 10^{04}$ | $0.218^{39}$ | 160 | $1.56 \times 10^{01}$ | $1.65 \times 10^{01}$ | $2.77 \times 10^{01}$ |
| 624 Hektor | $1.13 \times 10^{05}$ | $0.034^{40}$ | 122 | $1.22 \times 10^{02}$ | $1.29 \times 10^{02}$ | $2.16 \times 10^{02}$ |

**Table 4** Average sensitivity (SNR = 10, $R \sim 250$: $\lambda < 2.42~\mu m$, $R \sim 60$: $\lambda > 2.42~\mu m$) and saturation photon fluxes per-spectral bin for Twinkle's three spectrometers.

| Instrument property | Exposure time (s) | Number of observations | Flux (photons/m²/s) | | |
|---|---|---|---|---|---|
| | | | Ch0 | Ch1 | Ch2 |
| Saturation limit | 1 | 1 | $1.47 \times 10^{07}$ | $6.53 \times 10^{07}$ | $5.90 \times 10^{07}$ |
| Sensitivity limit | 300 | 1 | $3.89 \times 10^{01}$ | $4.08 \times 10^{01}$ | $4.01 \times 10^{02}$ |
| Sensitivity limit | 300 | 10 | $1.19 \times 10^{01}$ | $1.21 \times 10^{01}$ | $1.26 \times 10^{02}$ |

## 8 Appendix B: Sensitivity Limits for Signal-to-Noise Ratio = 100

Similarly to Figs. 10 and 11, Fig. 13 shows the photon flux from various Solar System objects as well as the sensitivity and saturation limits for various exposure times for an SNR > 100.

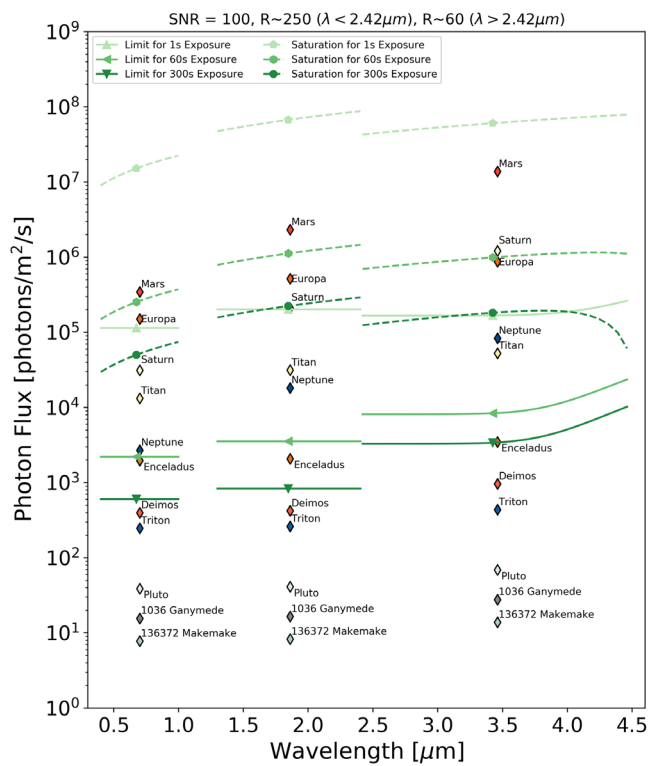

**Fig. 13** The average photon flux received per-spectral bin at Earth for various Solar System bodies at their average distance during observable periods with Twinkle. In addition, the sensitivity and saturation limits of Twinkle are plotted for various exposure times assuming that a SNR > 100 is required.

## 9 Appendix C: Availability of Dwarf Planets and Major Asteroids

Comparisons of the ephemerides of dwarf planets (Figure 14) and major asteroids (Figure 15) show long, periodic viewing windows for these bodies over the mission lifetime considered. Hence, while observers may have to wait for a suitable window, many opportunities for characterization will exist for each body.

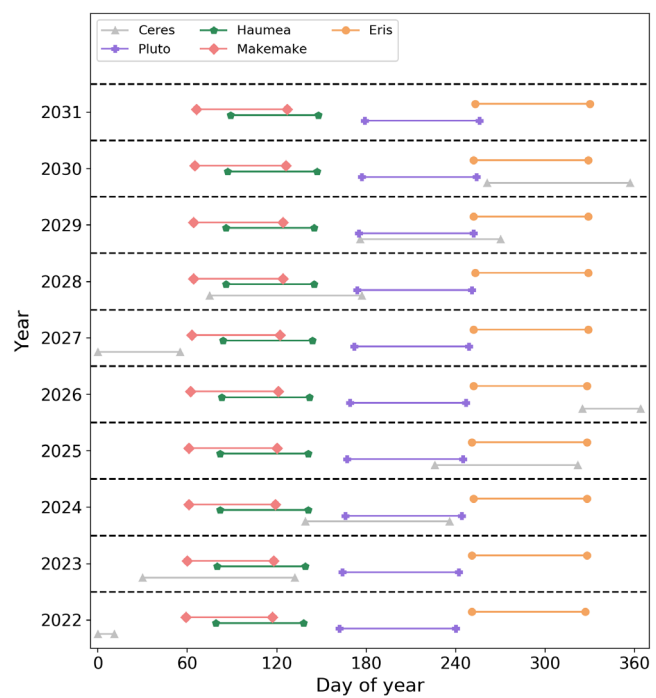

**Fig. 14** Period for which the dwarf planets lie within Twinkle's field of regard from January 1, 2022, to January 1, 2032.

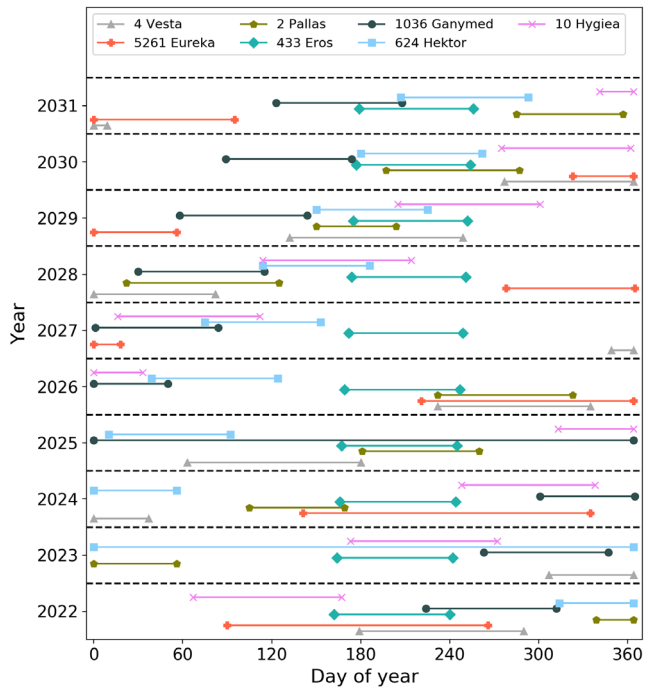

**Fig. 15** Period for which the major asteroids considered here lie within Twinkle's field of regard from January 1, 2022, to January 1, 2032.

## Acknowledgments

This work has been funded by the ERC Consolidator Grant ExoLights (GA 617119) and the STFC Grant Nos. ST/P000282/1, ST/P002153/1, and ST/S002634/1. We have used the JPL Horizons module of the astroquery python package. We thank the anonymous reviewers for their comments, which undoubtedly improved the quality of the manuscript.

**Billy Edwards** is a PhD student at University College London. His research focuses on capability studies of upcoming space-based telescopes for planetary and exoplanetary science. He is leading several studies into the science performance of Twinkle and is also heavily involved in the ESA M4 mission Ariel. He previously studied a master's degree in space science and engineering at UCL/MSSL and has a bachelor's degree in physics from the University of Bath.

**Giorgio Savini** has worked at UCL since 2009 on metamaterial concepts for satellite optics, and mid-/far-infrared modulation techniques for spectroscopy and interferometry. He worked on the testing of optical components and the ground-calibration campaign for the Planck satellite's HFI instrument as well as the software pipeline validation for the Spire spectrometer on the Herschel Space Observatory. He is the chief technology officer for Blue Skies Space Ltd. and payload lead for the Twinkle mission.

**Giovanna Tinetti** is a professor of astrophysics at University College London and director of the new UCL Centre for Space Exoplanet Data at Harwell. She is the principal investigator of Ariel, the European

Space Agency's next medium-class (M4) science mission. She is also cofounder and codirector of Blue Skies Space Ltd., which aims at creating new opportunities for science space satellites.

**Marcell Tessenyi** is the CEO of Blue Skies Space Ltd. and project manager for the Twinkle mission. He is responsible for the day-to-day programmatic activities of the Twinkle project. He has a PhD in astrophysics from the University College London in exoplanet spectroscopy. His contributions to space instruments include the European Space Agency's M3 candidate mission EChO and the M4 mission Ariel.

**Claudio Arena** is a PhD student at University College London and is part of the Astronomical Instrumentation Group. His research focus is on fine guidance systems (FGS). His research includes simulations of Twinkle's FGS performance, as well as modeling, simulating, and breadboard level testing a novel FGS design using piezoelectric actuators.

**Sean Lindsay** is a lecturer at the University of Tennessee whose primary research is on determining the mineralogy and relative abundances of dust species for the small bodies of the Solar System. He has developed various tools to reduce and analyse visible, near infrared, and thermal infrared spectroscopic data for a variety of instruments on ground and space-based observatories.

**Neil Bowles** is an associate professor at the University of Oxford. His main research interests are in laboratory measurements that help analyse and interpret data returned from space-based remote sensing and *in-situ* instruments for landers. He also works on developing new space-based instrumentation. He is a coinvestigator and science team member on numerous ESA and NASA missions, including Ariel and Mars Insight.